# Dynamics of rapidly spinning blob-filaments: fluid theory with a parallel kinetic extension

J. R. Myra,[1] J. Cheng[2] and S. E. Parker[2]

[1] *Lodestar Research Corporation, 360 Bobolink Court, Louisville, CO 80027, USA*
[2] *Department of Physics, University of Colorado, Boulder, CO 80309, USA*

**Abstract**

Blob-filaments (or simply 'blobs') are coherent structures formed by turbulence and sustained by nonlinear processes in the edge and scrape-off layer (SOL) of tokamaks and other magnetically confined plasmas. The dynamics of these blob-filaments, in particular their radial motion, can influence the scrape-off layer width and plasma interactions with both the divertor target and with the main chamber walls. Motivated by recent results from the XGC1 gyrokinetic simulation code reported on elsewhere [J. Cheng et al. submitted to Nucl. Fusion and available at https://arxiv.org/abs/2302.02877v1], a theory of rapidly spinning blob-filaments has been developed. The theory treats blob filaments in the closed flux surface region or the region that is disconnected from sheaths in the SOL. It extends previous work by treating blob spin, arising from partially or fully adiabatic electrons, as the leading order effect and retaining inertial (ion charge polarization) physics in next order. Spin helps to maintain blob coherency and affects the blob's propagation speed. Dipole charge polarization, treated perturbatively, gives rise to blob-filaments with relatively slow radial velocity, comparable to that observed in the simulations. The theory also treats the interaction of rapidly spinning blob-filaments with a zonal flow layer. It is shown analytically that the flow layer can act like a transport barrier for these structures. Finally parallel electron kinetic effects are incorporated into the theory. Various asymptotic parameter regimes are discussed and asymptotic expressions for the radial and poloidal motion of the blob-filaments are obtained.

Keywords: blob, filament, spin, adiabatic, electron kinetics, zonal flow

# I. Introduction

Plasma turbulent transport in the edge (i.e., closed flux surface region) and scrape-off layer (SOL) plasma plays a critical role in the operation of tokamaks and most other magnetic confinement devices. Turbulent transport in the edge influences edge plasma gradients of density and temperature, which indirectly help to determine the overall core confinement properties of the device. In the SOL, those plasma profile gradients set the width of the exhaust heat flux channel and hence the heat loads on the divertor target plate and its short term and long-term survivability. Plasma that can transport long distances across the SOL may also impact the main chamber walls where it can cause erosion, impurity release and recycling.

An important physical process associated with turbulence and transport in the edge and SOL is the intermittent generation by edge turbulence of coherent structures often called blobs or blob-filaments. In this paper the terms blob-filament, blob and filament will be used almost interchangeably, the choice depending on whether the particular emphasis of the discussion is on perpendicular (blob) or parallel (to B, filament) structure.

Blobs were observed by optical emission as far back as the 1980's in the Caltech tokamak.[1] Their importance for SOL transport and a theoretical mechanism for their structure and propagation was highlighted in a classic paper much later.[2] A review of blob theory[3] and another on the comparison of blob theory and experiment[4] describe progress as of 2008 and 2011 respectively. Considerable effort, both in theory, simulation and experiment have been devoted to the topic in the intervening years. Some of those papers that are most relevant to the present work will be cited subsequently.

A blob-filament is basically a magnetic flux tube that has excess density and pressure relative to the surrounding plasma. Rapid parallel plasma flow and/or transport elongates the structures in the direction of the magnetic field resulting in parallel scale lengths of many, even tens of, meters. In the direction across the magnetic field, blobs usually have dimensions somewhat comparable to the turbulence that created them, typically cm-scale in order of magnitude.

The basic mechanism giving rise to the radial motion of blob-filaments is that the oppositely directed curvature and grad-B drifts of electrons and ions in the blob result in a charge-polarization-induced electric field, roughly vertical or poloidal in the outboard edge of a tokamak. The $\mathbf{E}\times\mathbf{B}$ drift then propels the blobs in the outward radial direction. The curvature-grad-B-induced current is ultimately closed to form a complete circuit by other mechanisms such as cross-field ion polarization current or in the SOL by the sheath and wall currents.[3,4]

Soon after these basic mechanisms were understood, it was recognized that blob spin could rotate and mix the polarization charges and thereby impact the motion of the blobs.[5] This physics was described using relatively simple models treating spin for sheath-connect SOL blobs that have an internal electron temperature gradient. In this case, the spin arises from the sheath-induced blob electric field, i.e., from the radial blob profile gradient of $\Phi \sim 3T_e$ where $\Phi$ is the electrostatic potential, and $T_e$ the electron temperature. Subsequent work addressed the role of Boltzmann spinning.[6,7,8]

It was shown in Ref. 6 that parallel variation of density along the filament will set up a parallel varying Boltzmann-like potential along the field line, which causes blob spin. The spin was shown to affect the blob dynamics when the Boltzmann potential exceeded the dipole potential. Realistic tokamak magnetic geometry also causes parallel variation along the filament,[7] and an associated reduction in the radial blob velocity was observed in these fluid simulations. Tilting of the polarization electric field in the RZ plane due to the Boltzmann potential converts radial blob motion into poloidal motion.[8,9]

Previous theory and 2D fluid simulations in a slab model also addressed the question of how a blob interacts with a localized background sheared flow. Ref. 10 developed a theory-based estimate for the destruction of a non-spinning blob from a combination of shearing by the flow and subsequent diffusion. Fluid simulations[11] considered the effect of a biased limiter on SOL sheath-connected non-spinning density blobs. It was shown that a relatively large sheared flow destroys non-spinning blobs. The 2D fluid model for $T_e$-induced spin for sheath-connected blobs in the SOL was also used to study the interaction of those blobs with sheared flows.[5] Spinning blobs were shown to survive in the presence of weak sheared flows, and the relative sign of vorticity of the blob and flow region were shown to be important. Similar interactions of vortices and sheared flows were reported in experiments on RFX-mod[12,13] and in fluid simulations.[14]

Blob suppression by sheared flow was shown in reduced model fluid simulations with the SOLT code.[15] In addition to blob suppression, gyrofluid simulations have shown that blobs inside and outside a shear layer can merge, thereby exchanging particles and heat and then subsequently break up giving rise to a novel transport process.[16] The parallel variation of flows has also been shown to reduce blob radial motion.[17] In other theoretical work, blob creation has been linked to the interaction of radial streamers with sheared flows.[18]

Several experimental papers have reported on the interaction of blobs with sheared flows. Correlations between poloidal flows and quiet periods in edge turbulence were seen in experiments on NSTX.[19,20] In the EAST tokamak, suppression effects on blobs and the reduction of associated radial transport were observed in experiments in which the flow shear was increased by the application of lower hybrid waves.[21] Blob distortion and blob splitting was observed in the ASDEX Upgrade tokamak from sheared flows induced by radio-frequency waves.[22] Eddy tilting and breaking by sheared flows was also observed on TEXTOR.[23] In the TORPEX device, it was shown that blobs can extract energy from sheared flows.[24] The generation of blob-filaments and their interaction with background flows and with the turbulent spectrum was studied in experiments in the HL-2A tokamak.[25]

A comprehensive study of blob properties using the gas puff imaging diagnostic was reported for the NSTX device including L- and H-mode plasmas[26] where the effects of flow shear layers might be expected to be quite different. A suppression of blobs in H-mode plasmas was observed. Blob suppression in Ohmic H-modes was also reported in another study.[27]

Blobs with negative radial velocity have been observed in experiments[26,28] and also in XGC1 simulations.[29] We will show in this paper that a possible explanation for negative radial velocity blobs is the interaction of spinning blobs with a flow shear layer.

In this paper, the theory of blob dynamics in the presence of the Boltzmann potential is extended by developing a perturbation theory in which the leading order consists of a rapidly spinning blob-filament. The theory is motivated by recent results from the XGC1 gyrokinetic simulation code, reported on elsewhere.[29] The XGC1 simulations employed drift-kinetic electrons and it was found that the electron response was dominantly adiabatic resulting in the observed leading order blob spin. The dipole potential giving rise to blob propagation in the theory is obtained in next order. The technique additionally allows an analytical treatment of the interaction of these rapidly spinning blob-filaments with a zonal flow layer. The theory was originally developed to describe blobs that were seeded in the closed flux surface region in the XGC1 simulations. It is thus directly applicable to that region. The theory is also applicable to blobs in the SOL when they are in the inertial regime, and are therefore not sheath connected. A minor extension to the theory could incorporate the sheath conductivity, but that will not be demonstrated here in order to limit the complexity and number of possible parameter regimes.

The plan of our paper is as follows. The derivation of the leading order (monopole) and first order (dipole) potentials is presented in Sec. II when there are no background sheared flows.

The propagation of blob density is given in Sec. III including the **E**×**B** drift from the dipole potential as well as the direct effect of curvature and grad-B drifts. Section IV gives explicit results for the blob velocity from the fluid theory in several different asymptotic limits depending on the spin rate, the collisionality, the perturbed electron pressure and the role of ion inertia. In Sec. V, the effect of a sheared background flow on blob propagation is considered. A condition for a localized flow to act as a barrier to blob motion is derived, again in the context of rapid blob spin. In Sec. VI the expression for the effective dipole conductivity derived in the fluid model is generalized to a kinetic description. Finally, a summary and conclusions are presented in Sec. VII. Some detailed results of the calculation are given in an Appendix.

## II. Monopole and dipole potentials: no background flows

### *A. Basic equations*

The basic equations of the model are the vorticity equation for $\Phi$ including a phenomenological vorticity (spin) damping term $\mu$ to model dissipation from plasma viscosity or neutral friction[3,5]

$$\frac{\omega_{pi}^2}{4\pi\Omega_i^2}\left(\frac{\partial}{\partial t} + \mathbf{v}_E \cdot \nabla + \mu\right)\nabla_\perp^2 \Phi = \nabla_\| J_\| + \frac{2c}{B}\mathbf{b}\times\mathbf{\kappa}\cdot\nabla p \tag{1}$$

and a simplified electron drift kinetic equation which will be used to provide the parallel current

$$\frac{\partial f}{\partial t} + \mathbf{v}_E \cdot \nabla f + v_\| \nabla_\| f + \frac{e}{m_e}\nabla_\| \Phi \frac{\partial f}{\partial v_\|} = C\{f\}. \tag{2}$$

Here $\omega_{pi}$ and $\Omega_i$ are ion plasma and ion cyclotron frequencies, $\mathbf{v}_E = (c/B)\,\mathbf{b}\times\nabla\Phi$ is the **E** × **B** velocity, $J_\|$ is the parallel current density, $\kappa$ is the curvature, $p = p_e + p_i$ is the plasma pressure, $f(v_\|)$ is the electron distribution function, $m_e$ is the electron mass and C is the Coulomb collision operator. The subscripts $\|$ and $\perp$ refer to the parallel and perpendicular projections with respect to the background magnetic field $\mathbf{B} = \mathbf{b}B$. The only warm ion, finite $T_i$ term retained is the contribution to the pressure on the right-hand side (RHS) of Eq. (1). Gyrokinetic ion effects will be treated elsewhere.

A kinetic treatment of the parallel current is given in Sec. VI. Here, to present the basic outline of the theory, we take the parallel velocity moment of Eq. (2) and use a Krook model for the collision term, $C\{f\} \rightarrow -\nu$. This results in

$$\frac{\partial}{\partial t}(nu_{\parallel}) + \mathbf{v}_E \cdot \nabla(nu_{\parallel}) + \nabla_{\parallel}(nv_{te}^2) - \frac{ne}{m_e}\nabla_{\parallel}\Phi + \nu nu_{\parallel} = 0 \quad (3)$$

where n is the electron density, $u_{\parallel}$ is the parallel fluid velocity, and $v_{te}$ is the electron thermal speed.

In the stead-state limit for an isothermal plasma this gives

$$\nu J_{\parallel} + \mathbf{v}_E \cdot \nabla J_{\parallel} = \frac{ne^2}{m_e}\left(\frac{T_e}{e}\nabla_{\parallel}\ln n - \nabla_{\parallel}\Phi\right) \quad (4)$$

where $T_e$ is the electron temperature. For example, in the collisional limit $\nu \gg \mathbf{v}_E \cdot \nabla$ the result is

$$J_{\parallel} = \sigma_{\parallel}^B \left(\frac{T_e}{e}\nabla_{\parallel}\ln n - \nabla_{\parallel}\Phi\right) \quad (5)$$

where

$$\sigma_{\parallel}^B = \frac{\omega_{pe}^2}{4\pi\nu} = \frac{ne^2}{m_e \nu} \quad (6)$$

This is just the usual collisional conductivity if we identify $\nu = 0.51\nu_{ei}$ where the 0.51 is a Braginskii coefficient.

The goal of this section is to solve the vorticity equation to obtain the dipole potential using a linearization of Eq. (4) for $J_{\parallel}$. Having obtained the dipole potential, we then go on to determine the blob convection velocity in Sec. III, first without background flows, then in Sec. V with background flows.

### *B. Derivation*

In leading order, we assume the blob is a cylindrically symmetric rapidly spinning structure with spin frequency $\Omega_b$ given by the $\mathbf{E}\times\mathbf{B}$ rotation frequency for a blob with a potential hill of radius $\delta_b$

$$\Omega_b = \frac{v_{E0}}{r} = \frac{c}{rB}\frac{d\Phi_0}{dr} \sim -\frac{c_{se}\rho_{se}}{\delta_b^2}\frac{e\Phi_0}{T_e} \quad (7)$$

Here $c_{se} = (T_e/m_i)^{1/2}$ and $\rho_{se} = c_{se}/\Omega_i$. We work in blob cylindrical coordinates, where z is along the magnetic field which is also the blob-filament axis, r is measured from the center of the blob and θ is an azimuthal angle around the blob with

$$x = r\cos\theta$$
$$y = r\sin\theta \tag{8}$$

where x is radial, y is poloidal and $\mathbf{b} = \mathbf{e}_z$. The zero-order blob is taken to be circular in cross-section, $n_0 = n_0(r)$, $\Phi_0 = \Phi_0(r)$ and $J_{\|0} = J_{\|0}(r)$.

We solve the vorticity equation for the azimuthally symmetric zero order (m = 0) and first order dipole (m =1) component of the potential where

$$\Phi \propto e^{-im\theta + ik_\| z} \tag{9}$$

For the m = 0, leading order potential $\mathbf{v}_E(r)\cdot\nabla\nabla^2\Phi(r) = 0$. Since the blob convection velocity is assumed to be very small compared with the spin velocity, we can also neglect $\partial_t\nabla^2\Phi$. For simplicity we also neglect μ in this order. It will cause decay of the blob vorticity on a slow time scale. The m = 0 component of the vorticity equation is then just $\nabla_\| J_{\|0} = 0$ or from Eq. (5) noting that $\mathbf{v}_E(r)\cdot\nabla J_{\|0} = 0$

$$\nabla_\| J_{\|,0} = \sigma_\|^B \nabla_\|^2\left(\frac{T_e}{e}\ln n_0 - \Phi_0\right) = 0 \tag{10}$$

Since $\kappa = -\mathbf{e}_x/R$ and $\mathbf{b}\times\kappa\cdot\nabla p(r)$ is a pure dipole contribution it does not appear in this equation. Here the subscript 0 indicates the m = 0 (monopole) component which is also the leading order term.

Since $n_0$ and $\Phi_0$ only depend on r, it is convenient to take $n_0 = n_0(\Phi_0)$ and define an adiabaticity parameter

$$\alpha_{ad} = \frac{T_e}{n_0 e}\frac{dn_0}{d\Phi_0} \tag{11}$$

with $\alpha_{ad} = 1$ in the adiabatic limit, and $\alpha_{ad} = 0$ in the collisional reduced MHD limit (i.e., where the pressure term is neglected in Ohm's law and current is proportional to electric field).

In the simplified slab model considered here, $J_{\|,0}$ is treated as independent of z. These definitions result in a zero-order blob structure given by

$$\ln n_0 = \alpha_{ad}\frac{e\Phi_0}{T_e} \tag{12}$$

$$J_{\|,0} = -\sigma_\|^B \nabla_\| \Phi_0(1 - \alpha_{ad}) \tag{13}$$

For the m = 1 dipole contribution, expanding the vorticity equation in the small parameter $\delta_b/R$ where $\delta_b$ is the blob radius yields

$$\frac{\omega_{pi}^2}{4\pi\Omega_i^2}(\mathbf{v}_{E0}\cdot\nabla + \mu)\nabla_\perp^2\Phi_1 + \frac{\omega_{pi}^2}{4\pi\Omega_i^2}\mathbf{v}_{E1}\cdot\nabla\nabla_\perp^2\Phi_0 = \nabla_\|J_{\|,1} + \frac{2c}{B}\mathbf{b}\times\boldsymbol{\kappa}\cdot\nabla p_0 \quad (14)$$

The curvature term on the RHS is considered to be a first order correction in this ordering. A short digression is required to express $J_{\|,1}$ in terms of $\Phi_1$ for the fluid model. (As previously mentioned, the evaluation of the relationship between $J_{\|,1}$ and $\Phi_1$ for a more general kinetic model is the subject of Sec VI.)

The electron continuity equation in leading order is

$$\frac{\partial n}{\partial t} + \mathbf{v}_E\cdot\nabla n = \frac{1}{e}\nabla_\|J_\| \quad (15)$$

(The small curvature terms in the continuity equation are not needed here for the calculation of $J_{\|,1}$ because the leading order part of the $J_{\|,1}$ term in the vorticity equation is already ordered to be competitive.) Since the simulations in Ref. 29 that motived this work show slowly moving blobs, with propagation velocity much less that the spin velocity, we again drop the time derivative $\partial_t \ll \mathbf{v}_{E0}\cdot\nabla$. Linearizing about $n_0(r)$ and $\Phi_0(r)$, using

$$\mathbf{v}_{E1}\cdot\nabla n_0 = \frac{dn_0}{d\Phi_0}\mathbf{v}_{E1}\cdot\nabla\Phi_0 = -\frac{dn_0}{d\Phi_0}\mathbf{v}_{E0}\cdot\nabla\Phi_1 \quad (16)$$

and $\mathbf{v}_{E0}\cdot\nabla = -i\Omega_b$ results in

$$\Omega_b\alpha_{ad}n_0\frac{e\Phi_1}{T_e} - \Omega_b n_1 = \frac{k_\|}{e}J_{\|,1}. \quad (17)$$

To complete the calculation of the linearized conductivity, Eq. (4) is similarly linearized to obtain

$$\nu J_{\|,1} - i\Omega_b J_{\|,1} = ik_\|ev_{te}^2 n_1 - \frac{n_0 e^2}{m_e}ik_\|\Phi_1 \quad (18)$$

where we drop $\mathbf{v}_{E1}\cdot\nabla J_{\|,0} \propto \nabla\nabla_{\|}\Phi_0$ under the parallel eikonal assumption $\nabla_{\|}\ln\Phi_0 \ll k_{\|}$. Eliminating $n_1$ from the previous two equations gives the desired relationship between the linearized parallel current and potential

$$J_{\|,1} = -ik_{\|}\sigma_{\|,1}\Phi_1 \qquad (19)$$

where the effective linearized conductivity in the fluid model is

$$\sigma_{\|,1} = i\frac{n_0 e^2}{m_e}\frac{(1-\alpha_{ad})}{[m\Omega_b + i\nu - k_{\|}^2 v_{te}^2/(m\Omega_b)]}. \qquad (20)$$

Here, $m = 1$, the azimuthal mode number of the blob, has been explicitly reintroduced for later reference. (Noting that $\Omega_b$ in Eq. (23) arises from $\mathbf{v}_{E0}\cdot\nabla$ acting on first order quantities, specifically the $\exp(-im\theta)$ factor, this generalization is straightforward.)

There is no parallel current in the adiabatic limit $\alpha_{ad} = 1$. The collisional MHD limit is $\alpha_{ad} = 0$ with $\nu \gg \Omega_b$, $k_{\|}^2 v_{te}^2/\Omega_b$. In this limit $\sigma_{\|,1} \to \sigma_{\|}^B$. Other limits such as large spin ($\Omega_b \gg \nu$, $k_{\|}^2 v_{te}^2/\Omega_b$), and large $k_{\|}v_{te}$ (i.e., $k_{\|}^2 v_{te}^2/\Omega_b \gg \Omega_b, \nu$) will be treated in Sec. IV. Fluid theory cannot properly treat the case where the denominator in Eq. (20) is very small; this case requires the kinetic analysis of Sec VI since it implies wave-particle resonance.

Having obtained $\sigma_{\|,1}$ Eq. (19) can be used to eliminate $J_{\|,1}$ from the linearized vorticity equation, Eq. (14). This results in

$$L_1\Phi_1 - k_{\|}^2\sigma_{\|,1}\Phi_1 = \frac{2c}{B}\mathbf{b}\times\boldsymbol{\kappa}\cdot\nabla p_0 \qquad (21)$$

where the operator $L_1$ is defined by

$$L_1 = \frac{\omega_{pi}^2}{4\pi\Omega_i^2}\left[(-im\Omega_b + \mu)\nabla_{\perp}^2 + \frac{c}{B}\nabla(\nabla_{\perp}^2\Phi_0)\cdot\mathbf{b}\times\nabla\right] \qquad (22)$$

Eq. (21) has an inertial limit in which $L_1$ (the ion polarization drift and viscous current) dominates and a parallel conductivity limit in which $k_{\|}^2\sigma_{\|,1}$ dominates.

For purposes of estimation, the operator $L_1$ may be simplified by making some rough approximations. First, we argue that the first term is larger than the second, i.e., we expect that $\nabla_{\perp}^2$ acting on the dipole potential $\Phi_1$ will be larger than on the monopole potential $\Phi_0$ simply because the former has more structure. This is not expected to be a strong inequality, but is used here for analytical simplification. Thus, $L_1$ is approximated as

$$L_1 \approx \frac{\omega_{pi}^2}{4\pi\Omega_i^2}(-im\Omega_b + \mu)\nabla_\perp^2 \sim \frac{i\omega_{pi}^2}{4\pi\Omega_i^2}\frac{(m\Omega_b + i\mu)}{\delta_b^2} \qquad (23)$$

where the final form is a rough scaling estimate.

The RHS of Eq. (21) is made more explicit by considering a circular Gaussian blob of amplitude $\delta n_0$ on a constant background density $n_{00}$

$$n_0(r) = n_{00} + \delta n_0 \exp(-r^2/2\delta_b^2) \qquad (24)$$

Using $\mathbf{b}\times\kappa\cdot\nabla = -R^{-1}\partial/\partial y$ where $\mathbf{b} = \mathbf{e}_z$ and $\kappa = -\mathbf{e}_x/R$, performing the y derivative and looking near the blob center where $\exp(-r^2/2\delta_b^2) \to 1$, Eq. (21) takes the form

$$L_1\Phi_{10} - k_\parallel^2 \sigma_{\parallel,1}\Phi_{10} = i\frac{2c(T_e + T_i)}{RB}\frac{r}{\delta_b^2}\delta n_0 \qquad (25)$$

In this final form, the complex notation of Eq. (9) has been employed, i.e., $y = r \sin\theta = \text{Re}[i\, r \exp(-i\theta)]$, and $\Phi_{10}$ is the coefficient of $\exp(-im\theta)$ for $m = 1$. The physical potential is obtained by taking $\text{Re}\,[\Phi_{10}\exp(-i\theta)] = \text{Re}\,[\Phi_{10}(x-iy)/r]$. Thus $\text{Re}(\sigma_{\parallel,1})$ gives rise to $\text{Im}(\Phi_{10})$ and a physical potential that varies in the y direction, for an $\mathbf{E}\times\mathbf{B}$ drift of the blob in the x (radial) direction, while $\text{Im}(\sigma_{\parallel,1})$ results in a blob $\mathbf{E}\times\mathbf{B}$ drift in the poloidal direction.

## III. Blob propagation

So far, we have used the vorticity equation for a rapidly spinning blob to obtain an explicit equation, Eq. (25), for the dipole potential that gives rise to blob propagation. The propagation of the blob density itself is of course described by the continuity equation, which is treated in this section.

The electron form of the continuity equation is

$$\frac{\partial n}{\partial t} + \mathbf{v}_E \cdot \nabla n = \frac{2c}{eB}\mathbf{b}\times\kappa\cdot(T_e\nabla n - ne\nabla\Phi) + \frac{1}{e}\nabla_\parallel J_{\parallel e} \qquad (26)$$

where on the closed surfaces, $J_\parallel \approx J_{\parallel e}$. Eliminating the parallel current term using the vorticity equation, Eq. (1), results in the ion form of the continuity equation

$$\frac{\partial n}{\partial t} + \left(\mathbf{v}_E + \frac{2cT_i}{eB}\mathbf{b}\times\boldsymbol{\kappa}\right)\cdot\nabla n = -\frac{2cn}{B}\mathbf{b}\times\boldsymbol{\kappa}\cdot\nabla\Phi + \frac{\omega_{pi}^2}{4\pi e\Omega_i^2}\left(\frac{\partial}{\partial t} + \mathbf{v}_E\cdot\nabla + \mu\right)\nabla_\perp^2\Phi \qquad (27)$$
$$-\frac{1}{e}\nabla_\parallel J_{\parallel i}$$

To describe the motion of the blob structure as a whole, it is sufficient to retain only the leading order potential $\Phi_0$ on the RHS because these terms are already small from the curvature and the convective time derivative respectively. Similar remarks apply to the density on the LHS. Noting that $\Phi_0(r)$ is approximately independent of time and gives zero in the convective term therefore the last term may be dropped (in the absence of background sheared flows) and we have approximately

$$\frac{\partial n_0}{\partial t} + \left(\mathbf{v}_E + \frac{2cT_i}{eB}\mathbf{b}\times\boldsymbol{\kappa}\right)\cdot\nabla n_0 = -\frac{2cn_0}{B}\mathbf{b}\times\boldsymbol{\kappa}\cdot\nabla\Phi_0 - \frac{1}{e}\nabla_\parallel J_{\parallel i} \qquad (28)$$

Using $\Phi_0 = T_e \ln n_0 / (e\alpha_{ad})$, from Eq. (12) the final leading order form of ion continuity for a spinning blob is

$$\frac{\partial n_0}{\partial t} + \left(\mathbf{v}_E + \mathbf{v}_{b\kappa}\right)\cdot\nabla n_0 = -\frac{1}{e}\nabla_\parallel J_{\parallel i} \qquad (29)$$

where

$$\mathbf{v}_{b\kappa} = \frac{2c(T_i + T_e/\alpha_{ad})}{eB}\mathbf{b}\times\boldsymbol{\kappa} \qquad (30)$$

The bulk motion of the blob is controlled by the $\mathbf{E}\times\mathbf{B}$ and curvature drifts, where $\mathbf{v}_E = \mathbf{v}_{E0} + \mathbf{v}_{E1}$. The latter are often small and neglected in previous theories of blob propagation, but can be important in our rapidly spinning blob theory because the dipole-induced propagation $\mathbf{v}_{E1}$ is suppressed by the spin allowing $\mathbf{v}_{b\kappa}$ to be competitive. Note that the electrons add to the ion curvature and grad-B drift of the blob, but in a direction that is *opposite* to the usual electron drift. This happens through the electrostatic potential term on the RHS of Eq. (28) because of the Boltzmann electrons.

Thus, we can expect several separate effects to control the motion of a spinning blob:

    (i)    the trivial effect of a blob being carried along by any background $\mathbf{E}\times\mathbf{B}$ flows $\mathbf{v}_{E0}$;

(ii) radial and poloidal blob propagation from the first order potential $\Phi_1$ contributing to $\mathbf{v}_{E1}$;

(iii) curvature and grad-B drift effects in $\mathbf{v}_{b\kappa}$ which will move a blob in the vertical direction (down in the usual ion-grad-B drift lower single null tokamak configuration) and consequently off of flux surfaces.

(iv) additional effects not yet included when the background flows have flow shear. Those will be discussed separately in Sec. V.

This paper is primarily focused on (ii) and (iv), but all of these effects have been observed in XGC1 simulations.

## IV. Some asymptotic limits

Here we explore some asymptotic sub-limits of the present theory of rapidly spinning blobs. The fundamental assumption of the paper is that $\Phi_0 \gg \Phi_1$ i.e., that the monopole potential responsible for the blob spin always dominates the dipole potential responsible for blob propagation. Within this fundamental ordering, in the following we consider different relative orderings of the three characteristic frequencies $\nu$, $k_\| v_{te}$, $\Omega_b$ and the ratio that governs the importance of the inertial response relative to the parallel conductivity, namely $L_1/(k_\|^2 \sigma_{\|,1})$. Substituting for $\sigma_{\|,1}$ from Eq. (20) and $L_1$ from Eq. (23) the condition $L_1 \ll k_\|^2 \sigma_{\|,1}$ to be small can be rewritten as

$$\rho_{se}^2 \ll \delta_b^2 \frac{k_\|^2 v_{te}^2 (1-\alpha_{ad})}{(\Omega_b + i\mu)(\Omega_b + i\nu - k_\|^2 v_{te}^2 / \Omega_b)} \tag{31}$$

Subsections A, B and C that follow consider the case where Eq. (31) is satisfied; Subsection D treats the opposite limit.

### A. Collisional limit

The collisional limit pertains when $\nu \gg \Omega_b$, $k_\|^2 v_{te}^2 / \Omega_b$, as seen from Eq. (20) and when $L_1/(k_\|^2 \sigma_{\|,1}) \ll 1$. In this limit one obtains

$$\sigma_{\|,1} = \sigma_\|^B (1-\alpha_{ad}) \tag{32}$$

If $\alpha_{ad} = 0$ the result is the collisional reduced MHD limit, as expected. From Eq. (31) $L_1$ is negligible when

$$\rho_{se}^2 \left| \frac{(\Omega_b + i\mu)\nu}{k_\|^2 v_{te}^2} \right| \ll \delta_b^2 (1 - \alpha_{ad}) \tag{33}$$

Since we assumed $\Omega_b \nu/(k_\| v_{te})^2 \gg 1$, this can only be true for rather large blobs, and the window for this regime could be narrow.

If these conditions hold, the conductivity is real and positive resulting in outward radial blob convection from the dipole potential ($\Phi_{10} \propto -i$, $\Phi \propto -y$, and $v_{E1x} > 0$). This is as expected from standard blob theory where the sheath conductivity, also real and positive, plays the role of $\sigma_{\|,1}$ here.

Specifically, invoking $L_1/(k_\|^2 \sigma_{\|,1}) \ll 1$ in Eq. (25), and using Eqs. (6) and (32) for $\sigma_{\|,1}$

$$\Phi_{10} = -i \frac{\delta n_0}{n_0} \frac{2(T_e + T_i)\nu}{e\Omega_e(1-\alpha_{ad})} \frac{r}{k_\|^2 R \delta_b^2} \tag{34}$$

Here, and in the following, factor $\delta n_0/n_0 = \delta n_0/n_0(0) = \delta n_0/(n_{00}+\delta n_0)$. This factor describes the well-known effect of reduced dipole potential and blob velocity in the presence of a background density. The dipole potential is given by

$$\Phi_1 = \mathrm{Re}\,\Phi_{10} e^{-i\theta} = -\frac{\delta n_0}{n_0} \frac{2(T_e + T_i)\nu}{e\Omega_e(1-\alpha_{ad})} \frac{y}{k_\|^2 R \delta_b^2} \tag{35}$$

The radial $\mathbf{E}\times\mathbf{B}$ drift velocity for $\alpha_{ad} = 0$ is

$$v_{Ex} = -\frac{c}{B} \frac{\partial \Phi_1}{\partial y} = c_s \frac{2\nu}{\Omega_e} \frac{\rho_s}{k_\|^2 R \delta_b^2} \frac{\delta n_0}{n_0} \tag{36}$$

where $c_s = [(T_e+T_i)/m_i]^{1/2}$ and $\rho_s = c_s/\Omega_i$. This radial velocity result reproduces the scaling law in the so-called 'RX' or collisional regime of the two-region model.[30] In this collisional regime, the blob drifts outward with a velocity that increases with $\nu$ and with $\delta n/n$. For $\alpha_{ad} = 1$, Eq. (31) is violated and we must pass to the inertial regime discussed later.

### B. Large $k_\parallel v_{te}$ limit

Next consider the case $k_\parallel^2 v_{te}^2 / \Omega_b \gg \nu, \Omega_b$, in the denominator of Eq. (31), but still assume that $L_1/(k_\parallel^2 \sigma_{\parallel,1}) \ll 1$. The result is

$$\sigma_{\parallel,1} = -i\sigma_\parallel^B \frac{\Omega_b \nu}{k_\parallel^2 v_{te}^2}(1-\alpha_{ad}) \tag{37}$$

By assumption $\sigma_{\parallel,1} \ll \sigma_\parallel^B$ and $\sigma_{\parallel,1}$ is also independent of $\nu$. The condition $L_1/(k_\parallel^2 \sigma_{\parallel,1}) \ll 1$ from Eq. (31) implies that

$$\rho_s^2 \ll \delta_b^2 (1-\alpha_{ad}) \left| \frac{\Omega_b}{\Omega_b + i\mu} \right| \tag{38}$$

which is not difficult to satisfy. The conductivity is imaginary and positive (since $\Omega_b < 0$), $\Phi_{10}$ is real and negative resulting in poloidal blob motion ($\Phi_{10} \propto -1$, $\Phi \propto -x$ and $v_{E1y} < 0$). Recall that in the present analysis $\mathbf{B} = B\mathbf{e}_z$ therefore the positive y direction is $\mathbf{e}_y = \mathbf{b} \times \mathbf{e}_\psi$ where $\mathbf{e}_\psi$ points radially outward. Specifically, we have

$$\Phi_{10} = \frac{2 T_e c_s}{e \Omega_b R(1-\alpha_{ad})} \frac{\delta n_0}{n_0} \frac{\rho_s r}{\delta_b^2} \tag{39}$$

and the poloidal $\mathbf{E} \times \mathbf{B}$ drift velocity for $\alpha_{ad} = 0$ is

$$v_{Ey} = \frac{c}{B} \frac{\partial \Phi_1}{\partial x} = c_s \frac{2 c_{se} \rho_{se} \rho_s}{\Omega_b R \delta_b^2} \frac{\delta n_0}{n_0} \tag{40}$$

The inequality given by Eq. (38) as well as the perturbation expansion condition $\Phi_1 \ll \Phi_0$ limit the maximum $v_{Ey}$ that can occur.

### C. Large spin limit

The other limit for $\sigma_{\parallel,1}$ is that of large spin, $\Omega_b \gg k_\parallel v_{te}, \nu$. The designation 'large spin' is used somewhat loosely here since the entire theory presented in this paper is a large spin limit. The fundamental ordering of the paper is $\Phi_0 \gg \Phi_1$ which, for example, is quite different from $\Omega_b \gg \nu$ considered here.

The effective conductivity is

$$\sigma_{\parallel,1} = i\frac{ne^2(1-\alpha_{ad})}{m_e\Omega_b} \tag{41}$$

For $L_1/(k_\parallel^2\sigma_{\parallel,1}) \ll 1$ the complex potential is

$$\Phi_{10} = -\frac{\delta n_0}{n_0}\frac{2(T_e+T_i)\Omega_b}{e\Omega_e(1-\alpha_{ad})}\frac{r}{k_\parallel^2 R\delta_b^2} \tag{42}$$

and the resulting poloidal velocity for $\alpha_{ad} = 0$ is

$$v_{Ey} = \frac{c}{B}\frac{\partial\Phi_1}{\partial x} = -c_s\frac{2\Omega_b}{\Omega_e}\frac{\rho_s}{k_\parallel^2 R\delta_b^2}\frac{\delta n_0}{n_0} \tag{43}$$

where again we use $\Phi_1 = \text{Re}\,\Phi_{10}e^{-i\theta}$.

### *D. Inertial limit*

The inertial limit pertains when $L_1/(k_\parallel^2\sigma_{\parallel,1}) \gg 1$, i.e. when the inequality in Eq. (31) is violated. In this case we can make the rough estimates

$$\Phi_{10} \sim \frac{2\Omega_i(T_e+T_i)}{e(\Omega_b+i\mu)}\frac{\delta n_0}{n_0}\frac{r}{R} \tag{44}$$

$$v_{Ex} = -\frac{c}{B}\frac{\partial\Phi_1}{\partial y} \sim c_s\frac{2\Omega_i\rho_s\mu}{(\Omega_b^2+\mu^2)R}\frac{\delta n_0}{n_0} \tag{45}$$

$$v_{Ey} = \frac{c}{B}\frac{\partial\Phi_1}{\partial x} \sim c_s\frac{2\Omega_i\rho_s\Omega_b}{(\Omega_b^2+\mu^2)R}\frac{\delta n_0}{n_0} \tag{46}$$

This large spin inertial limit is different from the zero-spin inertial limit considered in many earlier studies.[3,4] Here the inertial term $L_1\Phi_1$ is linear in the dipole potential; in the zero-spin theory there is no $\Phi_0$ to linearize about, and the inertial term is nonlinear in the dipole potential. This results in a different velocity scaling. The inertial limit here is most similar to the limit $W > 1$ considered in Ref. 5, where $W$ in that reference was defined as the ratio of the spin-dependent part of the inertial term to the sheath conductivity for a sheath connected blob. It was shown there that for $W > 1$ the spin reduced the radial blob velocity and gave rise to a poloidal velocity. This trend is also evident in Eqs. (45) and (46).

## V. Effect of background zonal flow on blob motion

As briefly reviewed in the introduction, previous theoretical studies have shown how a blob or localized vortex interacts with a sheared flow.[5,11-14] These studies suggest that tearing apart of the blob as well as modifying its propagation are possible. Experimentally, blobs with negative radial velocity have been observed[26,28] In particular, see Fig. 7. of Ref. 26 and note that Fig. 2 for an H-mode plasma shows some blobs confined inside the separatrix and one that appears to 'bounce' off the separatrix. Negative blob velocities have also been observed in XGC1 simulations.[17,29,31]

We will show in this paper that a possible explanation for negative radial velocity blobs is the interaction of spinning blobs with a flow shear layer. We also investigate other types of interactions within the context of the present rapidly spinning blob theory.

A background $\mathbf{E}\times\mathbf{B}$ zonal flow $\mathbf{v}_{Ef}$ will of course contribute to the net blob velocity $\mathbf{v}_E$ in Eq. (30). But there is a more subtle effect that was observed in recent seeded blob simulations.[29] Although the zonal flow is poloidal and toroidal, the blob's *radial* motion is influenced by the zonal electrostatic potential. Considering the interactions in the plane perpendicular to $\mathbf{B}$, the vorticity is defined as $\varpi = \mathbf{b}\cdot\nabla\times\mathbf{v} \approx (c/B)\nabla_\perp^2 \Phi$ where $\Phi$ includes the blob spin as well as the background sheared zonal flow. Since $\nabla_\perp^2 \Phi$ may also be interpreted as a charge density, vorticity interactions are also related to the interaction of distributed charges.

The derivation takes place in the frame of the background flow at the blob center. In this frame the starting point is the steady state version of the vorticity equation

$$\frac{\omega_{pi}^2}{4\pi\Omega_i^2}\left(\mathbf{v}_E \cdot \nabla + \mu\right)\nabla_\perp^2 \Phi = \nabla_\| J_\| + \frac{2c}{B}\mathbf{b}\times\boldsymbol{\kappa}\cdot\nabla p \qquad (47)$$

where now $\Phi$ contains zero order pieces describing the blob spin $\Phi_0(r)$ and the background zonal flow $\Phi_{0f}(x)$ and we wish to solve for the perturbation term $\Phi_{1f}$ resulting from the interaction of these flows. Let the dipole term that we obtained previously, driven directly by $\mathbf{b}\times\boldsymbol{\kappa}\cdot\nabla p$, still be denoted by $\Phi_1$ and let the new flow dependent addition be $\Phi_{1f}$.

$$\Phi = \Phi_0(r) + \Phi_{0f}(x) + \Phi_1 + \Phi_{1f} \qquad (48)$$

Within the small flow large spin ordering $v_{E0f}/\delta_b \ll \Omega_b$ adopted here, the zero order flow terms may be dropped in the derivation of the linearized conductivity and $\sigma_{\|,1}$ is still given by Eq. (20), except for the choice of m that we will return to. Within this limit the perturbed vorticity equation is

$$(L_1 - \sigma_{\parallel,1} k_\parallel^2)(\Phi_1 + \Phi_{1f}) = \frac{2cT_e}{B} \mathbf{b} \times \mathbf{\kappa} \cdot \nabla n_0 - \frac{\omega_{pi}^2}{4\pi\Omega_i^2}(\mathbf{v}_{E0} \cdot \nabla \nabla_\perp^2 \Phi_{0f} + \mathbf{v}_{E0f} \cdot \nabla \nabla_\perp^2 \Phi_0) \quad (49)$$

where the definition of $L_1$ in Eq. (22) must now be generalized to

$$L_1 = \frac{\omega_{pi}^2}{4\pi\Omega_i^2}\left([(\mathbf{v}_{E0} + \mathbf{v}_{E0f}) \cdot \nabla + \mu]\nabla_\perp^2 + \frac{c}{B}\nabla[\nabla_\perp^2(\Phi_0 + \Phi_{0f})] \cdot \mathbf{b} \times \nabla\right) \quad (50)$$

and we have used the fact that $\mathbf{v}_{E0} \cdot \nabla \nabla_\perp^2 \Phi_0 = \mathbf{v}_{E0f} \cdot \nabla \nabla_\perp^2 \Phi_{0f} = 0$ because $\mathbf{b} \times \nabla\Phi \cdot \nabla$ annihilates any function that only varies in the same direction that $\Phi$ varies in. Since zero order spin terms are assumed to be dominant in this theory, products of zero order flow terms on the RHS of Eq. (49) have been dropped. Following the same reasoning, the zero order flow terms on the LHS, $\Phi_{0f}$ and $\mathbf{v}_{E0f}$, may be dropped with respect to $\Phi_0$ and $\mathbf{v}_{E0}$ although there is no need to do that explicitly in the manipulations that follow. We continue to use the subscript '1' for the perturbation; however, in this case the perturbation will contain both m = 1 and m = 2 terms.

Using Eq. (21), which defines $\Phi_1$ in the absence of background flow, and canceling these terms leaves an equation for the effect of the background flows on the perturbed blob potential

$$(L_1 - \sigma_{\parallel,1} k_\parallel^2)\Phi_{1f} = -\frac{\omega_{pi}^2}{4\pi\Omega_i^2}(\mathbf{v}_{E0} \cdot \nabla \nabla_\perp^2 \Phi_{0f} + \mathbf{v}_{E0f} \cdot \nabla \nabla_\perp^2 \Phi_0) \quad (51)$$

The terms on the RHS have a simple physical interpretation. The first term represents the charge polarizing force from the interaction of the zonal vorticity $\nabla_\perp^2 \Phi_{0f}$ with the blob electric field $\mathbf{v}_{E0}$. The second term represents the charge polarizing force from the interaction of the blob vorticity $\nabla_\perp^2 \Phi_0$ with the zonal electric field $\mathbf{v}_{E0f}$.

In the rest frame of the blob center, the background flow varies across the blob as $v_{E0f} \approx x\, \partial v_{E0f}/\partial x$ provided that $x \sim \delta_b \ll L_f$ where $L_f$ is the scale length of the flow. This is the easiest case to treat and will be dealt with first. In that limit $\mathbf{v}_{E0} \cdot \nabla \nabla_\perp^2 \Phi_{0f} \ll \mathbf{v}_{E0f} \cdot \nabla \nabla_\perp^2 \Phi_0$ and Eq. (51) reduces to

$$(L_1 - \sigma_{\parallel,1} k_{\parallel}^2)\Phi_{1f} = -\frac{\omega_{pi}^2 T_e}{4\pi\Omega_i^2 e\alpha_{ad}} v'_{0f}\, x\, \frac{\partial}{\partial y}\nabla_{\perp}^2 \ln n_0 \qquad (52)$$

where $\mathbf{v}_{E0f} \equiv v_{0f}\mathbf{e}_y$ and $'$ denotes $\partial/\partial x$.

In the central portion of the blob, $r \ll \delta_b$, and taking the small blob amplitude limit $\delta n_0/n_0 \ll 1$ for analytic simplification, we find for the Gaussian profile of Eq. (24)

$$\frac{\partial}{\partial y}\nabla_{\perp}^2 \ln n_0 = \frac{4y}{\delta_b^4}\frac{\delta n_0}{n_0} + \ldots \qquad (53)$$

Note that substituting Eq. (53) into Eq. (52) gives $\Phi \propto xy$, which has a quadrupole structure that results in quite different dynamical behavior than the familiar blob potential dipole. For example, in the collisional limit one obtains a contribution to the blob velocity from its interaction with the sheared flow as

$$\mathbf{v}_{1f} = \frac{\nu c_{se}\rho_{se}}{\Omega_i\Omega_e k_{\parallel}^2(1-\alpha_{ad})\alpha_{ad}}\frac{4}{\delta_b^4}\frac{\delta n_0}{n_0} v'_{0f}\,(y\mathbf{e}_y - x\mathbf{e}_x) \qquad (54)$$

Because x and y change sign across the blob center, the left and right sides of the blob move in opposite directions, as do the top and bottom parts. When the flow pattern compresses the blob in x, it tears the blob apart in y, and vice versa. Net radial motion of the blob (repulsion or attraction to shear layer) can occur when $v'_{0f}$ is a function of x.

The local radial velocity of a given part of the blob structure is proportional to $-x\, v'_{0f} = -v_{0f}$ which in turn is proportional to $-\partial\Phi/\partial x = E_{0fx}$ i.e. the radial electric field creating the flow. Referring back to the discussion of Eq. (51), this shows that the local motion of the blob in the radial direction behaves like a positive charge, i.e., it moves in the direction of the background (zonal) electric field, as previously demonstrated.[29]

When the condition $\delta_b \ll L_f$ does not hold, both terms on the RHS of Eq. (51) must be retained. It is straightforward to show that the resulting equation is

$$\begin{aligned}(L_1 - \sigma_{\parallel,1} k_{\parallel}^2)\Phi_{1f} &= \frac{\omega_{pi}^2 T_e}{4\pi\Omega_i^2 e\alpha_{ad}}\left(\frac{1}{r}\frac{d\ln n_0}{dr} y\, v''_{0f} - v_{0f}\frac{\partial}{\partial y}\nabla_{\perp}^2 \ln n_0\right) \\ &\sim \frac{\omega_{pi}^2 T_e v'_{0f} \ln n_0}{4\pi\Omega_i^2 e\alpha_{ad}\delta_b}\left(\frac{1}{L_f} + \frac{1}{\delta_b}\right)\end{aligned} \qquad (55)$$

where in the second form, for later use, a size and scaling estimate has been obtained by replacing blob gradients by $1/\delta_b$ and flow gradients ($\partial/\partial x$) by $1/L_f$. The resulting radial blob velocity is proportional to $\partial \Phi_{1f}/\partial y$ and hence $v_{1fx}$ is proportional to

$$\hat{v}_{1fx} \equiv \frac{\partial^2 \ln n_0}{\partial y^2} \frac{\partial^2}{\partial x^2} v_{0f} - v_{0f} \frac{\partial^2}{\partial y^2} \nabla_\perp^2 \ln n_0 \qquad (56)$$

with a positive constant of proportionality. For example, in the collisional limit the constant is the pre-factor in Eq. (54), $\nu c_{se} \rho_{se} / [\Omega_i \Omega_e k_\parallel^2 (1-\alpha_{ad})\alpha_{ad}]$.

An example of the contribution to the radial blob velocity, using Eq. (56) and the Gaussian profile of Eq. (24) for $n_0$, is shown in Figs. 1 and 2. In Fig. 1, a sample zonal flow potential is given in normalized units (see caption) by $\Phi_{0f} = \exp[-(x+2)/4] + 4 \exp[-x^2/2]$. The radial electric field is just the negative of $v_{0f}$. The zonal profiles have shapes which are qualitatively similar to what is normally observed near the separatrix where the radial electric field changes sign. Note that the blob is of modest amplitude, $\delta n/n_0 = 1$ and that it extends over a region where the zonal flow varies significantly. Hence, this case has $\delta_b \sim L_f$.

For the calculation illustrated in Fig. 2, the left-hand-side of Eq. (55) is assumed to be in a dissipative regime (dominated by either $\nu$ or $\mu$). The interaction depends on the location of the blob center and the interaction of the blob with the zonal flow extends over a range of order $x_0-1$ to $x_0+1$. At $x_0 = -3$, panel (a), the blob experiences a net negative velocity, i.e., one that repels it from the shear layer concentrated at $x = 0$ (the 'separatrix'). The interaction also compresses the blob radially since the very left-most part of the blob ($x \sim -4$) moves slightly to the right. In Fig. 2 (b), at $x_0 = -1$, the net radial velocity is positive and to the right. We speculate that this change of net sign with the blob location $x_0$ in Fig. 2 (b) relative to Fig. 2 (a) may encourage blob bifurcation.

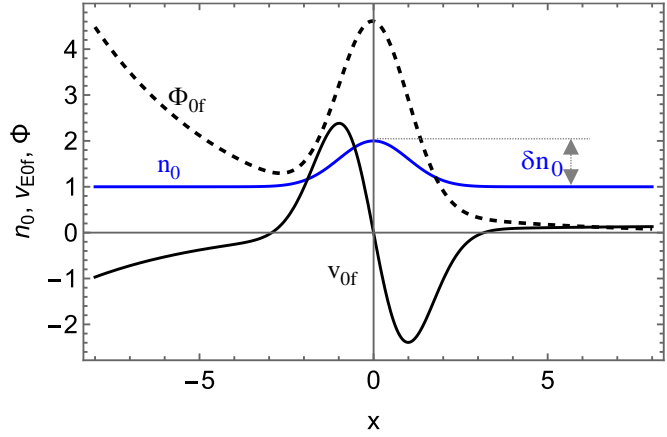

Fig. 1 Profile of the zonal flow potential $\Phi_{0f}$ (black dashed) and corresponding zonal velocity $v_{0f}$ (black) superimposed on the blob profile (blue) cut through its midplane, $y = 0$. The vertical scales are arbitrary, and the horizonal (spatial) scale is normalized to the blob radius. The background density $n_0$ and blob amplitude above background $\delta n_0$ are equal.

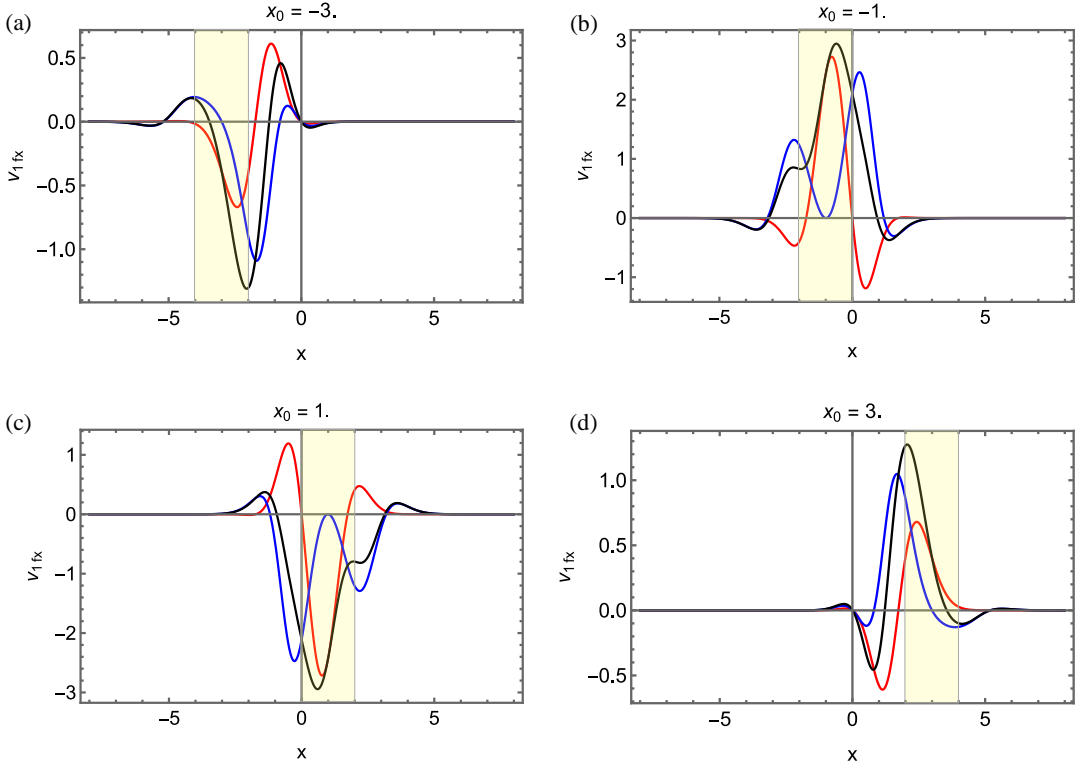

Fig. 2 Midplane cuts of the blob radial velocity $v_{1fx}$ from Eq. (56) for the zonal flow and blob density profiles illustrated in Fig. 1. The contribution of the first term from Eq. (56) is shown in red, the second term in blue and the total in black. The vertical scales are arbitrary, and the horizontal (spatial) scale is normalized to the blob radius. The position of the blob center $x_0$ is given in each panel and the approximate interaction zone, $x_0 \pm 1$, is indicated by the shaded region.

At $x_0 = +1$, Fig. 2 (c), the net radial velocity is again negative, which will tend to trap the blob near the separatrix, if the zonal flow is sufficiently strong. If the blob does escape further into the 'SOL' at $x_0 = +3$, Fig. 2 (d) shows that the zonal flow there contributes to its positive, right-going, velocity.

This illustration shows that a sheared flow layer can act as a barrier to radial blob propagation in the spinning blob regime, as seen in XGC1 simulations.[29] Since the radial blob velocity from the dipole (curvature and grad-B) interaction is outward, a necessary condition for a blob barrier (in addition to the direction shown in Fig. 2) is that the RHS of Eq. (55) exceeds the RHS of Eq. (21). Replacing blob gradients of these terms by $1/\delta_b$ and estimating $\kappa \sim -1/R$, we arrive at the approximate barrier condition for a rapidly spinning blob

$$C_b \equiv C_{b0} \frac{R v'_{0f}}{\Omega_i} \left( \frac{1}{L_f} + \frac{1}{\delta_b} \right) > 1 \qquad (57)$$

where $C_{b0}$ is an order unity numerical factor, roughly estimated as $C_{b0} \sim \ln n_0 / (2\alpha_{ad})$, but best obtained from simulations. From the simulations of Ref. 29 the deduced value obtained was $C_{b0} \approx 1/1.75 = 0.57$. When the condition $C_b > 1$ is met, the shear layer can repel the approaching side of the blob and prevent it from crossing the layer.

Within factors of order unity, it was shown that this condition was satisfied in the XGC1 simulations of Ref. 29 where both reflection and bifurcation were observed. Turning to experiments, taking the illustrative parameters for an Alcator C-Mod H-mode,[32] $v'_{E0f} = 2 \times 10^6$ /s, $\Omega_i = 2.4 \times 10^8$/s at $B = 5$ T, $R = 90$ cm and $L_f = \delta_b = 0.5$ cm we find $C_b/C_{b0} = 3.0$ indicating a substantial blob barrier.

The other effect of the quadrupole potential introduced by the sheared flow, evident in the $\delta_b \ll L_f$ limit, is the stretching of the blob in the direction orthogonal to the compression. This quadrupole distortion may also assist in the bifurcation of the blob.

Finally, note that the derivation of Eq. (57) makes no assumption about the regime the blob is in as long as it is rapidly spinning, $\Phi_0 > \Phi_1$. It could just as well apply to a sheath-connected spinning blob in the SOL. It is therefore interesting to compare Eq. (57) with the condition given by Yu and Krasheninnikov in Eq. (15) of Ref. 11 for a non-spinning sheath-connected blob. Their condition for significant interaction with a biasing potential may be written in the form $v'_{0f} > \Omega_i \delta_b L_f / (L_\| \rho_s)$ where $L_f \ll \delta_b$ has been assumed to insure a large-blob that would be in the sheath-connected limit and $L_\|$ is the magnetic field line connection length to the sheath. Comparing with Eq. (57) for $\alpha_{ad} \sim 1$ leads to the conclusion that a sheath-connected rapidly spinning blob will interact more strongly with a shear flow layer than a non-spinning blob if $L_\| \rho_s/(R\delta_b) < 1$. Of course, the interactions can be quite different in the two cases, since the spinning blob will tend to be more coherent as discussed next.

The interaction of a 'nonthermalized spinning blob' with a weak externally imposed shear layer was also considered in the sheath-connected limit in Ref. 5. Here the term 'nonthermalized spinning blob' refers to blob spin induced by a sheath potential $\Phi \sim 3T_e$ that varies radially across the blob because of its internal electron temperature gradient. Spin was shown in numerical studies to increase blob coherency and have different effects on the blob evolution depending on the relative sign of the blob vorticity and that of the shear layer.

## VI. Parallel electron kinetics

The goal of this section is to derive a kinetic expression for $\sigma_{\|,1}$ i.e., the relationship between the linearized parallel current and the parallel electric field, from a drift-kinetic theory which allows retaining the streaming term, $v_\| \nabla_\|$. The 1-D electron drift kinetic equation in the direction parallel (to B) in our model is

$$\frac{\partial f}{\partial t} + \mathbf{v}_E \cdot \nabla f + v_\| \nabla_\| f + \frac{e}{m_e} \nabla_\| \Phi \frac{\partial f}{\partial v_\|} = C\{f\} \tag{58}$$

where $C\{f\}$ is the electron-ion collision operator and the small curvature and grad-B drift terms are neglected since we will only use this equation to get the leading order relationship between the electrostatic potential and the parallel current. This should be valid under the easily satisfied assumption that the curvature and grad-B drift frequencies obey $\omega_\kappa$, $\omega_{\nabla B} \ll \Omega_b$ where as previously $\Omega_b$ is the blob spin rate.

In lowest order we take $f = f_0$ to be a Maxwellian

$$f_0 = \frac{n_0(\Phi_0)}{(2\pi)^{1/2} v_{te}} \exp\left(-\frac{v_\|^2}{2v_{te}^2}\right) \tag{59}$$

Since the blob moves slowly compare to the spin rate, the time derivative in Eq. (58) may be neglected. Furthermore $\mathbf{v}_{E0} \cdot \nabla f_0 = 0$ and $C\{f_0\} = 0$. In the adiabatic limit $f_0$ is a function of energy so that the terms involving $v_\parallel \nabla_\parallel f_0$ and $\partial f_0/\partial v_\parallel \nabla_\parallel \Phi_0$ cancel. More generally, in this section as previously, we simply assume for analytical expediency that $\nabla_\parallel$ on equilibrium quantities may be neglected. Then $f_0$ satisfies Eq. (58).

The equation for the linearized distribution function $f_1$ is

$$\frac{\partial f_1}{\partial t} + \mathbf{v}_{E0} \cdot \nabla f_1 + v_\parallel \nabla_\parallel f_1 - C\{f_1\} = -\mathbf{v}_{E1} \cdot \nabla f_0 - \frac{e}{m_e} \nabla_\parallel \Phi_1 \frac{\partial f_0}{\partial v_\parallel} \quad (60)$$

Decomposing into $f_1$ and $\Phi_1$ into Fourier components as before, anticipating $m = 1$ for the azimuthal variation

$$\Phi_1, f_1 \propto e^{-i\theta + ik_\parallel z} \quad (61)$$

yields

$$-i\Omega_b f_1 + ik_\parallel v_\parallel f_1 - C\{f_1\} = -\mathbf{v}_{E1} \cdot \nabla f_0 + \frac{ik_\parallel e \Phi_1}{T_e} v_\parallel f_0 \quad (62)$$

In this section, we do not consider background $\mathbf{E} \times \mathbf{B}$ flows, or equivalently, work in the flow frame. The time derivative has again been neglected compared with the spin rate.

To proceed, we use a Krook model for the collision term, $C\{f\} \to -\nu$ and note that

$$\nabla f_0 = f_0 \frac{\nabla n_0}{n_0} = \alpha_{ad} f_0 \frac{e}{T_e} \nabla \Phi_0 \quad (63)$$

Furthermore using $-\mathbf{v}_{E1} \cdot \nabla \Phi_0 = \mathbf{v}_{E0} \cdot \nabla \Phi_1 = -i\Omega_b \Phi_1$ Eq. (62) becomes

$$-i\Omega_b f_1 + ik_\parallel v_\parallel f_1 + \nu f_1 = \frac{e\Phi_1}{T_e}\left(-i\Omega_b \alpha_{ad} f_0 + ik_\parallel v_\parallel f_0\right) \quad (64)$$

In order to make an analogy to the usual perturbed 1D Vlasov equation in a homogeneous plasma it is convenient to temporarily define $\omega \equiv \Omega_b + i\nu$. This results in

$$f_1 = \frac{K_0 f_0 + K_1 v_\parallel f_0}{-i(\omega - k_\parallel v_\parallel)} \quad (65)$$

where

$$K_0 = -i\Omega_b \alpha_{ad} \frac{e\Phi_1}{T_e} \tag{66}$$

$$K_1 = ik_\| \frac{e\Phi_1}{T_e} \tag{67}$$

Applying $-e\int dv_\| v_\|$ one obtains

$$J_{\|,1} = -ie \int_{-\infty}^{\infty} dv_\| v_\| \frac{K_0 + K_1 v_\|}{\omega - k_\| v_\|} f_0 \tag{68}$$

Defining $\zeta = \omega/(2^{1/2} k_\| v_{te})$ and performing some algebraic manipulations, $J_{\|,1}$ can be cast into the form

$$J_{\|,1} = -i\sigma_{\|,1} k_\| \Phi_1 \tag{69}$$

where the linearized conductivity in our kinetic model is given by

$$\sigma_{\|,1} = -i \frac{n_0 e^2}{k_\|^2 T_e} [\Omega_b(1 - \alpha_{ad}) + i\nu][1 + \zeta Z(\zeta)] \tag{70}$$

and $Z(\zeta)$ is the plasma dispersion function. In obtaining this result, it is useful to note the following:

$$Z(\zeta) = \frac{1}{\pi^{1/2}} \int_{-\infty}^{\infty} dt \frac{e^{-t^2}}{t - \zeta}, \quad \text{Im } \zeta > 0 \tag{71}$$

$$\frac{1}{\pi^{1/2}} \int dt \frac{t e^{-t^2}}{\zeta - t} = -1 - \zeta Z(\zeta) \tag{72}$$

$$\frac{1}{\zeta \pi^{1/2}} \int dt \frac{t^2 e^{-t^2}}{\zeta - t} = -1 - \zeta Z(\zeta) \tag{73}$$

The integrals in Eqs. (72) and (73) are conveniently evaluated by adding and subtracting $\zeta$ in the numerator of Eq. (72) and $\zeta^2$ in the numerator of Eq. (73) to create a term that cancels the resonant denominator.

The conductivity $\sigma_{\parallel,1}$ derived here may be used directly in the expressions given in Sec. II, specifically in Eq. (25) in place of the fluid limit conductivity, to obtain the dipole potential $\Phi_1$ and the radial and poloidal blob velocities. The same replacement may be made throughout Sec. IV. Detailed unified results are given in Appendix A.

Before closing this section, we consider some asymptotic limits. The fluid limit is recovered by first invoking the large argument limit of the Z-function

$$1 + \zeta Z(\zeta) = -\frac{1}{2\zeta^2} - \frac{3}{4\zeta^4} + ..., \quad |\zeta| \gg 1 \tag{74}$$

to obtain

$$\sigma_{\parallel,1} = \frac{n_0 e^2}{m_e} \frac{i\Omega_b(1-\alpha_{ad}) - \nu}{(\Omega_b + i\nu)^2} \tag{75}$$

In the collisional sublimit $\nu \gg \Omega_b$ one recovers the collisional result of fluid theory, Eq. (32). In the collisionless sublimit, or large spin limit $\nu \ll \Omega_b$ the result agrees with Eq. (41).

In the small argument limit, the Z-function expands to

$$1 + \zeta Z(\zeta) = 1 + i\pi^{1/2}\zeta + ..., \quad |\zeta| \ll 1 \tag{76}$$

and the corresponding conductivity is

$$\sigma_{\parallel,1} = \frac{n_0 e^2}{k_\parallel^2 T_e}[-i\Omega_b(1-\alpha_{ad}) + \nu] \tag{77}$$

It suggests the possibility of a regime where the $\text{Re}(\sigma_{\parallel,1})$ increases with $\nu$ and hence the radial blob velocity *decreases* with $\nu$, opposite to the dependence in the collisional regime. However, the small argument limit of $Z(\zeta)$ is of questionable validity unless the plasma is also collisionless. This is because the Krook model probably cannot correctly capture long mean free path physics, $\nu \ll k_\parallel v_{te}$. In the more justifiable collisionless sublimit $\nu \ll \Omega_b$ one recovers Eq. (37), noting that it is independent of $\nu$ when $\sigma_\parallel^B$ is eliminated.

$$\sigma_{\parallel,1} = -i\frac{n_0 e^2}{k_\parallel^2 T_e}\Omega_b(1-\alpha_{ad}) \tag{78}$$

In general, the linearized parallel conductivity in our kinetic theory contains both real and imaginary parts. As a result, the dipole potential gives rise to both radial and poloidal blob motion. In particular, Eq. (70) implies finite Re $(\sigma_{\parallel,1})$ and consequently radial blob propagation in the regime $\zeta \sim 1$ as a result of electron Landau damping.

## VII. Summary and conclusions

In this paper, we have presented a theory of blob dynamics in a large spin ordering. The basic perturbation parameter is $\Phi_1/\Phi_0$ where $\Phi_0$ is the internal radial potential of the blob responsible for its spin and $\Phi_1$ is the dipole (or quadrupole) potential giving rise to radial and poloidal **E**×**B** blob propagation relative to the background flow.

The main results of our paper are Eq. (25) for $\Phi_1$ where $L_1$ is given by Eqs. (22) or its estimate in Eq. (23) and the effective plasma conductivity $\sigma_{\parallel,1}$ in the fluid model is given by Eq. (20). In the parallel electron kinetic generalization of the fluid model $\sigma_{\parallel,1}$ is given by Eq. (70). It was shown that the kinetic result reproduces the fluid result in appropriate limits. A minor extension of the theory could also include the sheath conductivity in $\sigma_{\parallel,1}$ making the results applicable to sheath-connected blobs in the SOL.

The total velocity of the blob is given as the sum of **E**×**B** and curvature drifts in Eqs. (29) and (30) where **E** includes the background zonal flows and the gradients of the dipole (or quadrupole) blob potential. Explicit expressions for $\Phi_1$ and the resulting radial and poloidal propagation velocities of the blob are given in various asymptotic limits in Sec. IV. Results are qualitatively in agreement with XGC1 simulations in order of magnitude. Detailed comparisons will be presented elsewhere.

Special attention was given to the interaction of the blob with a zonal flow shear layer. The flow-induced potential on the blob in the small flow ordering, $v_{E0f}/\delta_b \ll \Omega_b$ (or $v'_{E0f} < \Omega_b$) is given in Eq. (51). It was demonstrated that a spinning blob can be reflected by a sufficiently strong shear layer and a rough analytical criterion for this transport barrier effect was given in Eq. (57). It was shown that the criterion was satisfied in recent XGC1 seeded blob simulations where the transport barrier was observed, and the barrier estimate was shown to be relevant to typical Alcator C-Mod H-mode plasmas. Another important result from the shear layer analysis is the generation of a quadrupole blob potential, as shown in Eq. (52). The quadrupole potential should encourage blob bifurcation within a shear layer through both radial dynamics illustrated in Fig. 2 and by poloidal stretching combined with radial compression, as seen in Eq. (54).

The present calculation has two major deficiencies: (i) the Krook collisional model does not allow a rigorous treatment of collisions, especially in the long mean free path limit; and (ii) the calculation does not take into account the parallel variation of the zero-order potential, or of other zero order equilibrium quantities such as the magnetic field geometry. A parallel variation of the first order dipole potential is introduced through $k_\parallel$. This is only rigorous within the model if it is driven by the parallel variation of the curvature, which is regarded as first order. In reality, other effects such as the magnetic or $\mathbf{E}\times\mathbf{B}$ drift of the entire filament off a field line or instability of the filament can give rise to parallel variation. Thus, the model presented here is idealized in many respects. However, these idealizations allow analytical estimates for the magnitudes of the blob velocity and the transport barrier criterion.

Initial qualitative and some quantitative comparisons of the theory with both experiments and simulations are promising.[29] These include the observed coherency and longevity of nearly circular blobs, blob bifurcation near the separatrix, negative radial motion of blobs, and a shear layer barrier for blob propagation. If future work continues to show acceptable agreement of the theory with simulations and experiments, it will add to the available tools for understanding edge and scrape-off layer turbulence and transport.

## Acknowledgements

This research was supported by the SciDAC-4 project High-fidelity Boundary Plasma Simulation funded by the U.S. Department of Energy, Office of Science, Office of Fusion Energy Sciences under contract number DE-AC02–09CH11466. Discussions with members of the HBPS SciDAC project and with D.A. Russell are gratefully acknowledged.

## Appendix A: Unified expressions for the dipole-induced blob velocity

Recall that in the present analysis $\mathbf{B} = B\mathbf{e}_z$ therefore the positive y direction is $\mathbf{e}_y = \mathbf{b} \times \mathbf{e}_\psi$ where $\mathbf{e}_\psi$ points radially outward, in the positive x direction. With this convention the scaling of the rapidly spinning dipole-induced blob velocity can be summarized in a convenient form by collecting together suitable dimensionless combinations.

The starting point is Eq. (25) with $L_1$ estimated by Eq. (23) and m = 1. After a small amount of algebra, the blob velocity from the dipole electric field can be written as

$$v_{Ex} = -c_{se} \frac{\rho_{se}}{R} \operatorname{Im} \frac{S}{D} \qquad (A1)$$

$$v_{Ey} = c_{se} \frac{\rho_{se}}{R} \operatorname{Re} \frac{S}{D} \qquad (A2)$$

$$S = \frac{2\omega_{pi}^2}{\Omega_i \Omega_b}\left(1 + \frac{T_i}{T_e}\right)\frac{\delta n_0}{n_0} \qquad (A3)$$

$$D = L + i\sigma \qquad (A4)$$

$$\Omega_b = \frac{c_{se}\rho_{se}}{\delta_b^2} \qquad (A5)$$

$$L = \frac{\omega_{pi}^2}{\Omega_i^2}\left(1 + \frac{i\mu}{\Omega_b}\right) \qquad (A6)$$

$$\sigma = \frac{4\pi k_\parallel^2 \delta_b^2}{\Omega_b} \sigma_{\parallel,1} \qquad (A7)$$

$$\nu = 0.51 \nu_{ei} \qquad (A8)$$

Throughout, perpendicular gradients of the blob have been estimated as $\nabla_\perp \to 1/\delta_b$. These results are therefore to be interpreted as rough order of magnitudes estimates of the blob velocity that may also be useful for determining how the blob velocity scales with various parameters. Either the fluid form for $\sigma_{\parallel,1}$ given by Eq. (20) or the parallel electron kinetic form in Eq. (70) may be employed in Eq. (A7).

A compact result may be written as

$$v_{Ey} - iv_{Ex} = c_s \frac{\rho_s}{R} \frac{2\varepsilon_\perp \Omega_i}{\varepsilon_\perp(\Omega_b + i\mu) + 4\pi i k_\parallel^2 \delta_b^2 \sigma_{\parallel,1}} \frac{\delta n}{n} \qquad (A9)$$

where $\varepsilon_\perp = \omega_{pi}^2/\Omega_i^2$ is the low frequency plasma dielectric. Substituting for $\sigma_{\parallel,1}$ explicitly using the fluid limit form gives

$$v_{Ey} - iv_{Ex} = c_s \frac{\rho_s}{R} \frac{\delta n}{n} \frac{2\varepsilon_\perp \Omega_i (\Omega_b + i\nu - k_\parallel^2 v_{te}^2 / \Omega_b)}{\varepsilon_\perp (\Omega_b + i\mu)(\Omega_b + i\nu - k_\parallel^2 v_{te}^2 / \Omega_b) - k_\parallel^2 \delta_b^2 \omega_{pe}^2 (1 - \alpha_{ad})} \quad (A10)$$

from which all of the asymptotic limits of Sec. IV may be recovered. The expressions in this appendix do not include the interactions of the blob with a sheared flow. They are just the fundamental dipole curvature and grad-B induced propagation velocities.